\pdfoutput=1
\documentclass[11pt,onesided]{article}
\usepackage{natbib}
\DeclareMathDelimiter{(}{\mathopen} {operators}{"28}{largesymbols}{"00}
\DeclareMathDelimiter{)}{\mathclose}{operators}{"29}{largesymbols}{"01}
\usepackage{booktabs}
\usepackage[justification=centering]{caption}
\usepackage[flushleft]{threeparttable}
\usepackage{csquotes} % For quotes
\usepackage{amsmath} % assumes amsmath package installed
\usepackage{amssymb}  % assumes amsmath package installed
\usepackage{epsfig} % for postscript graphics files
\usepackage{mathptmx} % assumes new font selection scheme installed
\usepackage{adjustbox}
\usepackage{times} % assumes new font selection scheme installed
\usepackage[ruled]{algorithm2e} % For algorithms
\usepackage{multirow}
\usepackage{url}
\usepackage{listings}
\usepackage{lipsum}

\usepackage{lmodern}
\SetAlFnt{\small}
\SetAlCapFnt{\small}
\SetAlCapNameFnt{\small}
\SetAlCapHSkip{0pt}
\IncMargin{-\parindent}

\usepackage{amsthm}
\theoremstyle{definition}
\newtheorem*{definition*}{Definition}
\usepackage{booktabs} % For formal tables
\usepackage{balance}
\date{}

\begin{document}
% Title portion
\title{Content based News Recommendation via Shortest Entity Distance over Knowledge Graphs} 

\author{Kevin Joseph \\ KevinJoseph774@gmail.com \and  Hui Jiang \\ hj@cse.yorku.ca}

\maketitle

\begin{abstract}
	Content-based news recommendation systems need to recommend news articles based on the topics and content of articles without using user specific information. Many news articles describe the occurrence of specific events and named entities including people, places or objects. In this paper, we propose a graph traversal algorithm as well as a novel weighting scheme for cold-start content based news recommendation utilizing these named entities. Seeking to create a higher degree of user-specific relevance, our algorithm computes the shortest distance between named entities, across news articles, over a large knowledge graph. Moreover, we have created a new human annotated data set for evaluating content based news recommendation systems. Experimental results show our method is suitable to tackle the hard cold-start problem and it produces stronger Pearson correlation to human similarity scores than other cold-start methods. Our method is also complementary and a combination with the conventional cold-start recommendation methods may yield significant performance gains. The dataset, CNRec, is available at: https://github.com/kevinj22/CNRec
\end{abstract}

\section{INTRODUCTION}

News recommendation is a pivotal part of many major web portals and it is important for these search portals to recommend the \enquote{right} news articles, at the right time. Recommending news articles is one of the most challenging recommendation tasks because news providers generally allow users to read news articles anonymously. As a consequence, most news recommendation systems have to cope with the hard \enquote{cold start} problem: making recommendations without explicit user feedback.

A major difficulty in content-based news recommendation is that the similarity between news articles does not necessarily reflect their relatedness. For instance, two news articles might share a majority of words, yet their actual topic could be very different. Additionally one user's opinion of what is a good recommendation may significantly differ from another's. Over-specialization is also a problem with news recommendation \cite{webist14}. In the news domain, many articles on the same topic are typically written differently on separate news portals. These articles, if recommended to the same reader, may result in a poor recommendation. Furthermore, the unstructured format of news articles makes it more difficult to automatically analyze their content than structured properties such as goods or services. 

\begin{figure}
	\centering
	\includegraphics[scale=0.15]{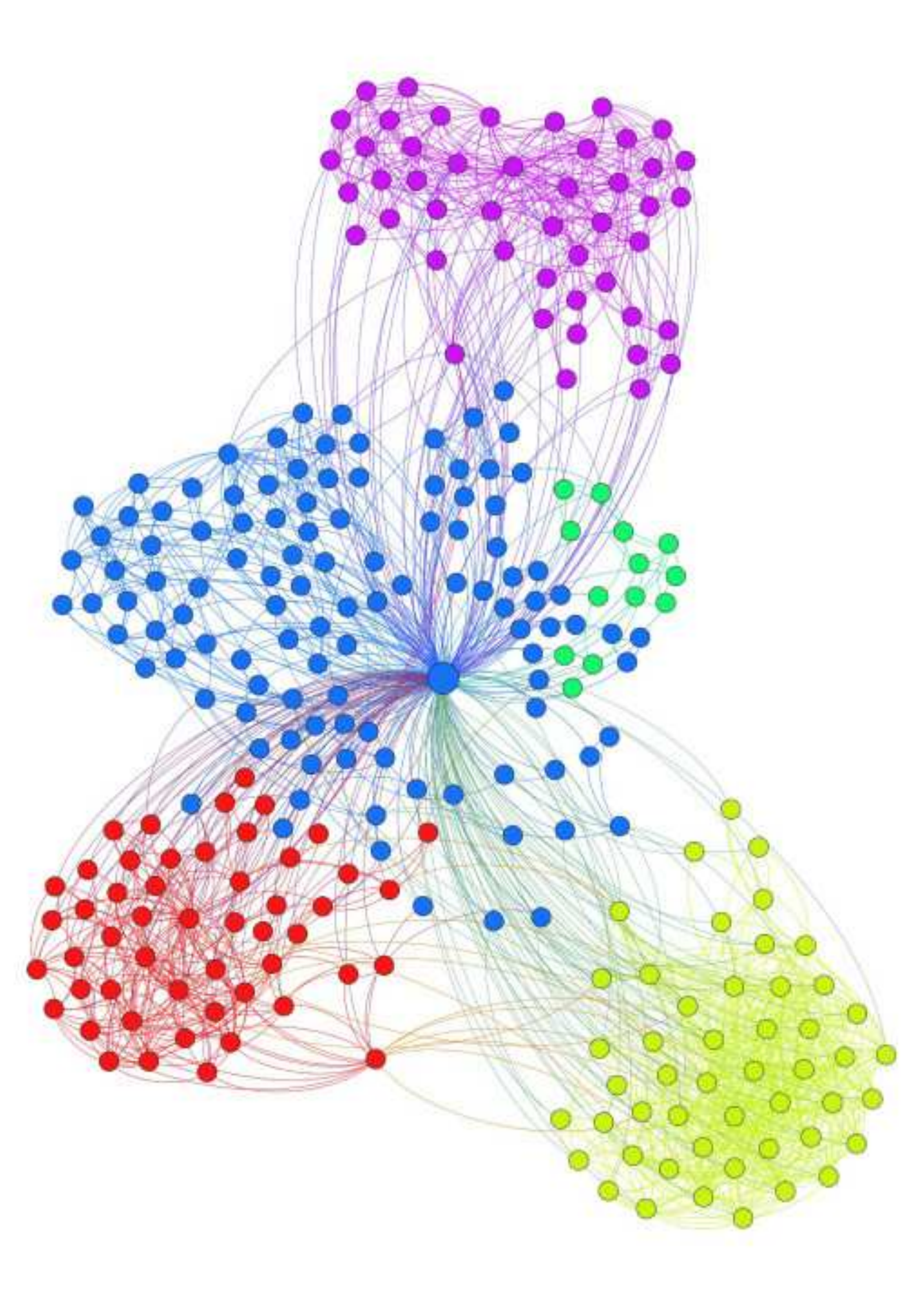}
	\caption{Each news article is represented as a subgraph (with different colours) in a large knowledge graph via named entity disambiguation.}
\label{Fig-articles-over-KGs}
\end{figure}

A common element across news stories is the description of specific events and/or updates about particular entities including people, places or objects. As these entities already represent key data related to the article, any additional information extracted about them could be used to make more informed recommendations. 
For instance, two news articles might share many common words, but the extra information gained through a knowledge graph may help to differentiate them. 
Through named entity disambiguation (NED), we determine the proper entity link from an optimized knowledge graph. Based on these entity links, 
a sub-graph can be created for each article, as shown in Figure \ref{Fig-articles-over-KGs}.
These articles are effectively embedded in the same space over the knowledge graph. 

If the sub-graphs do not completely overlap, we assume these two articles mention related topics, but not the exactly same ones. In this case, it may be a good idea to make a content-based recommendation based on their distance over the knowledge graph. 
To do this we propose a new content-based news recommendation algorithm, which leverages the vast amount of information between these sub-graphs from a curated knowledge graph.  
This algorithm, similar to \cite{KGD}, computes the single source shortest path distance 
(SSSP), between two sub-graphs. Our experimental results have shown our proposed method produces stronger Pearson correlation to human similarity scores than other cold-start recommendation methods. Our method is also complementary to these conventional cold-start recommendation methods since an ensemble of them may yield significant performance gains. As the knowledge graph used in this work, Freebase \cite{freebase:datadumps}, is extremely large, we focus our work on attaining the best results using the least data. 

 Unfortunately the similarity of two documents does not necessarily indicate that they would be a good recommendation to a user.  Thus bench-marking using the LeePincombeWelsh (LG50) \cite{LG50} similarity dataset such as in \cite{EfficientGraph}, \cite{LDSD}, and \cite{KGD} would not be a good indicator of recommendation performance. 

 To get direct feedback, live recommendation tests as in \cite{LiveTesting} can be used, although this is time consuming and expensive. Live tests also exhibit time constraints that may not allow parameter tuning with algorithms requiring pre training or pre computations. For the reasons above we believe it is necessary to provide a new human annotated data set, {\it CNRec}. {\it CNRec} provides human annotated document-wise similarity scores as well as whether a pair of articles is considered as a good recommendation by human. This paper will also cover the creation of  {\it CNRec} \footnote{CNRec is available here: https://github.com/kevinj22/CNRec.}. All results in this paper utilize {\it CNRec}. We hope {\it CNRec} will be used as a benchmark task for evaluating various cold-start news recommendation tasks in the research community. 

\section{Related Work}

Here we will briefly review related work on two topics: graph-based document similarity and content based news recommendation.

\subsection{Graph-Based Document Similarity}

{\em PathSim} in \cite{PathSim} defines a path between two sub-graphes A and B as the connectivity via the number of paths from A to B divided by the total number of path instances from A to B and B to A. {\em HeteSim} in \cite{HeteSim} computes the relatedness between A and B as the overlap between the out-neighbors of A and the in-neighbors of B. Leal et al. \cite{DbLeal} use a proximity metric as a measure of similarity among nodes, where proximity is defined as the state of being near in space, time, or relationship. Lam et al. \cite{DbLam} use a TF-IDF-inspired edge weighting scheme, combined with Markov centrality, to compute entity similarity. Palma et al. \cite{AnnSim} describe an annotation similarity measure, called {\em AnnSim}, to evaluate similarity using a 1-1 maximal matching of annotations. {\em SemStim} algorithm \cite{SemStim} is based on PageRank and completes many random walks over all neighbors of a nodes. 

Nunes et al. \cite{SCS} present a document similarity approach where a document-wise connectivity score is computed based on the number of paths between document annotations. In a follow-up paper 
,Nunes et al. use traditional TF-IDF to select top entities for each document. In \cite{LDSD,DbMusic}, Passant also computes all paths between two nodes and the number of direct and distinct links between resources in a graph, which are used to determine the similarity of two entities for recommendation. 

Zhu {\it et. al.} \cite{7572993} propose to use {\em WordNet} \cite{Miller:1995:WLD:219717.219748} and {\em DBpedia} \cite{Auer:2007:DNW:1785162.1785216}
to determine the semantic similarity between concepts. They have attained the state of the art performance using a weighted shortest distance metric based on the least common subsumer measure extracted from graphs. This method measures similarity between pairs of nodes in a graph. In comparison, our work compares news articles by groups of nodes, i.e. sub-graphs. 

In \cite{KGD}, similarity is measured based on entity linking and analysis of entity neighborhood in {\em DBpedia}, where information content is used to weight edges to compute the SSSP between all pairs entities in documents. In this work, we use a much larger knowledge graph, i.e., Freebase \cite{freebase:datadumps}, and compute the average minimum symmetric distance across all pairwise entities between two articles.

\subsection{Content-based News Recommendation}

Term frequency inverse document frequency ({\em TF-IDF}) is a simple and popular method originally proposed for information retrieval \cite{Ramos2003UsingTT}. TF-IDF is also widely used to determine the relative frequency of words in documents for content-based recommendation \cite{webist14}. 

{\it Doc2vec} in \cite{DBLP:journals/corr/LeM14} is an extension to learn an embedding vector for each document, where the order of words in the document is ignored and the objective is to predict a context word given the concatenated document and word vectors \cite{DBLP:journals/corr/LauB16}. The learned document embedding vectors can be simply used to compute similarity between documents for content-based recommendation. 

Bayesian Networks is a popular algorithm to model user interests in news recommendation \cite{a4e1b868beda4b46a6723ba3d024ec41}, \cite{Wen:2012:HAP:2109236.2109445}. 
Context trees for news recommendation are proposed in \cite{Garcin:2013:PNR:2507157.2507166} 
LDA is used in \cite{Li:2011:PNR:2336266.2336268} to represent topic distributions \cite{webist14}. 
\cite{Das:2007:GNP:1242572.1242610} proposes locality sensitive hashing (LSH)  and \cite{Li:2011:PNR:2336266.2336268} uses {\em MinHash}, i.e., a probabilistic clustering method. {\em MinHash} has also been combined with probabilistic latent semantic indexing (pLSI) in \cite{Das:2007:GNP:1242572.1242610}. 

In \cite{Tavakolifard:2013:TNP:2487788.2487930}, a pre-trained named entity recognition (NER)model is applied to each news article. Based on the identified entities from articles, they determine user's long-term preferences for collaborative filtering. In this paper, rather than building user profile, we also leverage named entity information to improve content based recommendation rather than collaborative filtering.
\cite{Li:2011:PNR:2336266.2336268}, \cite{Tavakolifard:2013:TNP:2487788.2487930} and \cite{webist14} survey more natural language processing techniques used for news recommendation. 

\subsection{Collaborative Filtering}

In collaborative filtering, recommendations are done by using the other preferences which are similar to the current users' past preferences. As this work focuses solely on content based recommendations this will not be covered here.

\section{Data Set: {\it CNRec} }

{\it CNRec} provides document to document similarity as well as whether a pair of articles was considered a good recommendation. The {\it CNRec} data set consists of 2700 pairs of news articles, selected from 30 groupings of 10 articles of human perceived similarity \ref{CNrec_Group_Ex}. In total we have 300 unique news articles originally published in a period of 3 consecutive days between August 25-28, 2014. Each article is paired with all other articles in the same group.  This results in 45 pairs that should produce positive similarity ratings. Another 45 pairs are randomly generated across other groups resulting in 2700 total pairs. The 3 day period, as well as the grouping and pairing procedure, provides a ideal set of articles to process. It allows engineers to focus on direct algorithm design rather than filtering relevant articles by time, or overlapping entities \cite{EfficientGraph}, before computations. 

Each pair of articles is rated by 6 human annotators against two questions:

\begin{enumerate}
	\item In terms of content delivered, how similar do you think these two articles are? The annoators were given 3 choices: {\it Not Similar / Similar/ Very Similar}. Their answers were converted into numerical values 0/1/2 .
	\item If one of these articles was recommended based on the other would you have followed the link? Each annotator choose between {\it NO} and {\it YES}, which were converted to numerical values 0/1.
\end{enumerate} 

\begin{figure}[h]
	\centering
	\includegraphics[scale=0.185]{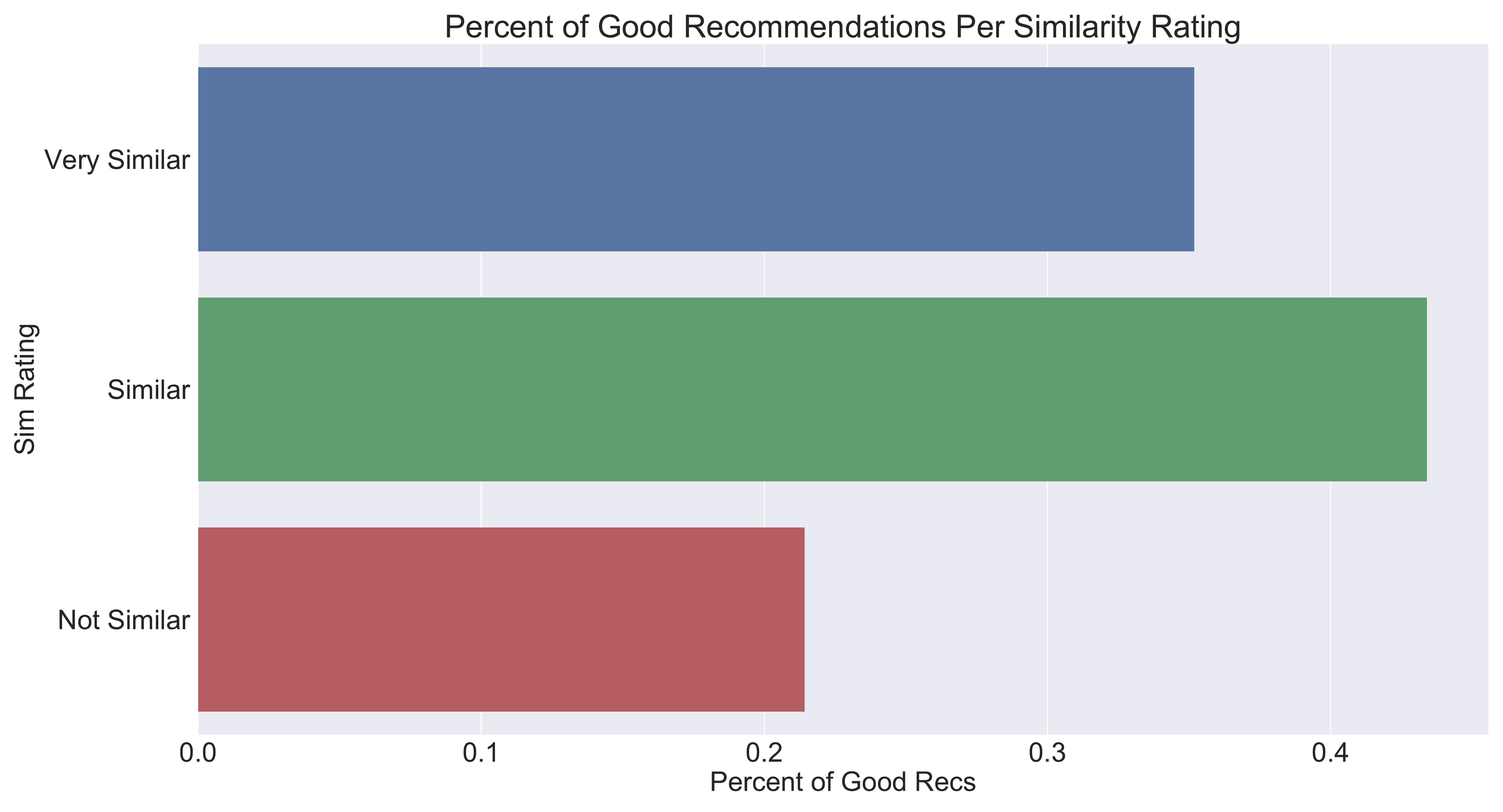}
	\caption{Similarity Rating Percent of Good Recommendations} 
	\label{Sim_Vs_Good}
\end{figure}

Referencing Figure \ref{Sim_Vs_Good}, {\it CnRec} indicates there is a 21.42\%, 43.40\%, 35.18\% split between similarity ratings leading to a good recommendation and a 73.71 Pearson correlation between similarity rating and good recommendations. This indicates that a high similarity does not mean a good recommendation. 

\begin{figure}[h]
	\begin{enumerate}
		\item D'oh! Homer Simpson takes ALS Ice Bucket Challenge
		\item McHendry Local pastor takes part in Ice Bucket Challenge
		\item Pub manager to have ice cold water poured over him tonight
		\item Woman has different take on ALS Ice Bucket Challenge
		\item Let me tell you about my hero, the man who beat ALS
		\item Facebook user claims that only 7\% of ALS challenge donations go towards charity
		\item Anti-Abortion Activists Have a Big Problem With the Ice Bucket Challenge
		\item ALS Ice Bucket Challenge Charity challenge hits Chelmsford
		\item Downton Abbey's Michelle Dockery's completes the ALS Ice Bucket challenge
		\item George RR Martin does Ice Bucket Challenge (then slips in the pool)
	\end{enumerate}
	\caption{Example CNRec Similar Group Article Titles}
	\label{CNrec_Group_Ex}
	\vspace{-4mm} 
\end{figure}

\section{Entity Shortest Distance Over Knowledge Graphs}

In this section, we introduce the main ideas of the method we propose for content-based news recommendation, namely entity shortest distance over knowledge graphs (SED). 

\subsection{Knowledge Graphs}

Many public Knowledge Graphs (KGs) are currently available. These include Freebase, DBpedia, and YAGO. Each network records millions of concepts, entities and their relationships. 

As news articles are centered on various named entities, one could hypothesize that by using a knowledge graph composed of said entities, a reasonable recommendation could be made. 
In this paper all experiments are conducted  with the Freebase \cite{freebase:datadumps} KG. In Freebase all relationships are modeled in the Resource Description Framework (RDF) format \cite{freebase:datadumps}, where each node edge pair is modeled in an object-orientated manner as object-predicate-subject. A leaf, a vertex of out degree 1, in Freebase is a human readable string used to assign values to an entities, such as birthdays or genders. The edges in Freebase are labeled by a predicate which exhibits a {\it domain.type.property} relationship between its related vertices \cite{freebase:datadumps}. Note that Freebase is originally modeled as a directed graph but we do not consider the direction of edges because most relations have a sound inverse relation, such as {\it nationalyOf} and {\it ofNationality}. This decision follows that of \cite{KGD}, \cite{SemStim}, and \cite{7572993}. 

Freebase is a huge knowledge graph consisting of approximately 1.9 billion triplets \cite{freebase:datadumps}. Any undirected simple graph G holds $|E| <= ( |V| * (|V|-1) ) / 2$ edges. This implies $|V|^2$ quadratic growth in the number of edges as |V| increases. Thus the number of edge traversals required to find the shortest path grows exponentially along with path length. Therefore we optimize the original Freebase KG by pruning some redundant and uninformative vertices and edges. In particular, all non-English entries are removed ( {\it CnRec} is English only). Many nodes in Freebase with out-degree less than 20 are generally filled with redundant information, thus they are also removed in our experiments. Nodes that appeared too frequency in our initial path searches and their associated edges were removed. For example {\it Male}, {\it people.person},  {\it olympic\_participating\_country}, and {\it United States}; these all fail to provide any specific information for recommendation. In this way, we may retain the most prominent information, remove noise, and reduce computational complexity. The English only Freebase KG is composed of 438 000 million triplets, while the completely pruned Freebase has 17.5 million triplets.

\subsection{The Proposed SED Algorithm}

Our proposed SED recommendation algorithm consists of the following steps: 
\begin{enumerate}
\item Entity discovery and linking: identify all named entities from each news article and link each entity to a node in Freebase
\item Sub-graph generation: create a sub-graph from the knowledge graph based on the linked nodes
\item Shortest distance similarity between articles: compute pair-wise shortest distance between any two documents over a union of their corresponding sub-graphs. 
\end{enumerate}

\subsubsection{Entity Discovery and Linking (EDL): mapping news articles to KG}

Given each news article we can use entity discovery and linking (EDL) to map it to an knowledge graph. EDL first identifies all named or nominal mentions of entities from unstructured text data. Next it classifies them onto proper types: person (PER), location (LOC), Organization (ORG), Geopolitical Entity (GPE) and Facility (FAC). Lastly, EDL needs to link each identified entity to a correct entity node in a knowledge graph. For example, in a sentence like ``{\it Apple just released its new iPhone.}'', ``{\it Apple}'' needs to be a node in Freebase associated with {\it Apple Inc.} rather than the fruit. In this work, we do not study how to improve EDL performance, but directly use an off-the-shelf EDL system \cite{iflytek:edl:2016,yorkU:edl:2016}. This system won the TAC KBP trilingual EDL competition in 2016 \cite{tac2016overview}.  

\subsubsection{Sub-graph Creation}

After EDL, each news article may be represented as a sub-graph in the shared knowledge graph. To produce a sub-graph, we start with a set of discovered entities N and create a labeled undirected graph $G=(V,E)$. First we set of nodes V of G be composed of all discovered entities: $V = N$. Next for each entity n in N we expand the nodes in G. This is done by performing a breadth first search on Freebase adding all outgoing relations. Loops are not added in graph creation. When used with our proposed weighting scheme all connections between node pairs are collapsed into one. All related predicates are then stored in an edge list. As our proposed weighting scheme considers the number of common neighbors, these paths would all be weighted equally. This removes the necessity of keeping all unique connections between nodes, reducing computational complexity. We expand up to a maximal length L for each discovered entity effectively adding all simple paths of up to length L.  From this we attain a sub-graph of Freebase. This graph is composed of the discovered entities and their related concepts along all paths of maximal length L that connect them. 

\subsubsection{Shortest Distance Between Articles}

The most intuitive information for similarity or relatedness between any two articles is the distance between their sub-graphs in the KG. Conceptually speaking,  the closer these two sub-graphs are located to each other, the more similar these two articles may be. 

Let $\mathcal{P}(n_i, n_j) = \{ P_1,P_2,...,P_k \}$ be the set of all possible paths connecting the nodes $n_i$ and $n_j$ with cardinality or size N. Let $|P_i|$ denote the length of a path $P_i \in \mathcal{P}(n_i, n_j) $, then $D(n_i, n_j) = \min(|P_k|)$ denotes the shortest path length between two nodes, namely shortest distance between them over $G$.

With EDL, a set of named entities is extracted from each article. After linking them to nodes in KG, each article is modeled by a subgraph in KG. As symmetric algorithms are provided in \cite{PathSim}, \cite{HeteSim}, \cite{GED}, and \cite{EfficientGraph} and one can easily derive asymmetric version we only present the symmetric version of SED. We measure the similarity between two articles, e.g. $S_1$ and $S_2$, based on the shortest distance between their subgraphs over the KG, denoted $\mathcal{D}(S_1, S_2)$ as:

$$\mathcal{\hat{D}}(S_1,S_2) = \frac{\mathcal{D}(S_1,S_2)  + \mathcal{D}(S_2,S_1)  }{2}$$
where
$\mathcal{D}(S_1,S_2) = \frac{ \sum_{m \in S_1 }  \min_{ n \in S_2}   D( m, n)} { |S_1|}$.

Note that this method provides an average minimum row-wise distance. This can easily be modified to use the average or sum of distance across all all pairwise nodes between two articles. 
By using the minimum we focus on the best pair of entities which has a higher probability of cooccurring in documents. This follows \cite{EfficientGraph} which computes an average maximum symmetric similarity between two articles. If we impose weights onto all edges in KG, we may compute weighted shortest distance as the sum of all weights along the traversed edges.

\section{Creating the Optimal Subgraph}

There are many practical issues to be considered in creating the subgraph for each news article. In this section, we will discuss some of these practical issues. 

\subsection{Maximal Length Expansion L}

Does a larger graph have a significant impact on performance? Following a large body of evidence from work \cite{UnsupGraphTopic}, \cite{KGD}, \cite{PathSim} and \cite{EfficientGraph} we only consider 1 and 2 hop graph creation. Additionally by using the union of two articles subgraphs, these 1 and 2 hops per node, which have been noted to have high predictive power, allow for longer relevant path lengths between the articles. 

\subsection{Pre-Screening Entities}
\label{subsection-PreScreen}

\begin{figure}[h]
	\centering
	\includegraphics[scale=0.185]{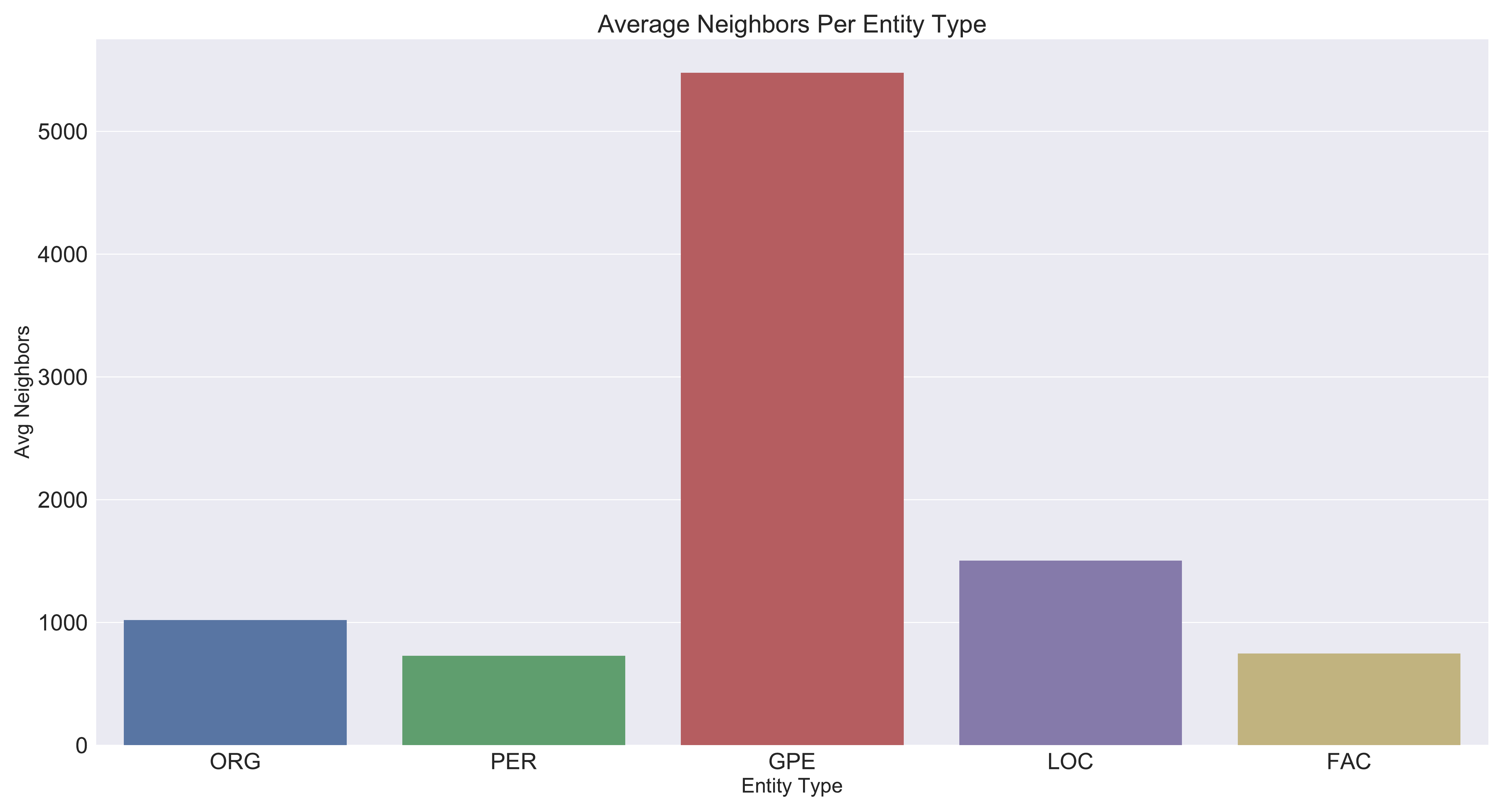}
	\caption{Average Entities Per Type} 
	\label{Ent_Per_Type}
	\vspace{-4mm} 
\end{figure}

Assume a set of named entities is identified by the EDL system \cite{xu-jiang-watcharawittayakul:2017:Long} for an article, how do we know whether some entities are more important than others or to treat them all equally?  One may perceive that the more an entity appears in an article, and the earlier it appears, the more closely related it is to the main theme of the article. Combining these two statistics, we rank all entities by a stable sort: first in descending order by count then in ascending order by offset. Over {\it CnRec} 11 entities are found on average per article. In addition only a single article returns 0 entities.  The EDL system also returns an entity type for each identified entity, such as Person (PER), Location (LOC), Organization (ORG), Geopolitical Entity (GPE) and Facility (FAC). Some types have many more neighbouring nodes in the KG than others, such as GPE as shown in Figure \ref{Ent_Per_Type}. Others such as LOC are frequently mentioned in articles. To insure there is no over saturation of common entities we can filter out entities. For our experiments we remove LOC and GPE typed entities. This leaves 7.5 entities on average per article. We then try filtering the top 5 and 8 entities per article. 

\subsection{Augmenting Context Words}

In this work, we use an off-the-shelf EDL system,\cite{tac2016overview} that is not designed to capture all Freebase entities. Moreover, some keywords in a news article may not be an above-defined entity at all. For example, in an article about Ebola, a debilitating virus, the word {\it Ebola} may be a key concept to find related articles. {\it Ebola} is not a named entity by definition. To address these issues, we propose to augment some \enquote{context words} from each news article. We use TF-IDF to score words for each article taking the top scoring $N$ words which exactly match a Freebase node's title. These $N$ words are added to the article's sub-graph. This method insures no articles have 0 discovered entities. In this paper, we experiment with $N=2-4$ context words per article. 

\subsection{Penalty for Disconnected Entities}

If there is no path between two entities across a pair of articles setting a `penalty` value could impact the recommendation quality. In the event the penalty value is set too low, too many articles would be recommended. If it is set to high, not enough articles will be recommended. In this paper we experiment within a range of penalty values. 

\section{Weighting Graph Edges}

In this work, we propose to weigh each Freebase edge based on social network influence weighting \cite{SocialInfluence}, which is named {\it relation weighting scheme (RWS)} hereafter.

\subsection{Relation Weighting Scheme}
\label{subsection-RWC}

The proposed relation weighting scheme (RWS) is based on social network influence weighting \cite{SocialInfluence}. Given a sub-graph $G$ for an article, assume we have an edge, $e$, which is connected to two entity nodes,  i.e., $n_i$ and $n_j$. For each entity, we may query the KG and obtain all first-order neighbouring nodes of each to form two neighborhood sets, denoted as $\mathcal{N}_i$ and $\mathcal{N}_j$.  To weigh the edge, $e$, between $n_i$ and $n_j$,  we first compute the conditional probability as follows:

$$ p(n_j | n_i) = \frac{| \mathcal{N}_i \cap \mathcal{N}_j |}{ |  \mathcal{N}_i|}. $$

If two nodes share many neighbours in common it is likely that they are closely related. As we are computing the shortest distance, a lower edge weight is better. As we are querying a pair of neighboring nodes $p(n_j | n_i) > 0$, then a weight, $w_i$, is assigned to $e$ as:
$w_i = 1 - p(n_j | n_i)$.

\subsection{Other Weighting Schemes}

In \cite{KGD}, and \cite{Esraa} many metrics for ranking edges in a KG are proposed. In this work, we implement JointIC and \cite{Esraa}'s top three  metrics to weight edges for our news recommendation tasks: i)  AF: normalized Attribute Frequency of the attributes over KG, favors common attributes in KG; ii) IAF: inverse attribute frequency for the attributes over KG, favors rare attributes; iii) AF-IAF: multiply the previous two features. 

\section{Experiments}

In this section, we evaluate the recommendation performance of the proposed methods. These are evaluated on a newly-created news recommendation benchmark data set,{\it CNRec}, on some typical cold-start settings. The proposed method is compared with conventional approaches used for cold-start content-based news recommendation, including TF-IDF and doc-vector. It is also compared with other graph traversal algorithms in terms of $F_1$, precision, recall, Pearson, and Spearman correlation with human labeled scores. All experiments with SED use the pruned version of FreeBase in which we removed nodes with out degree less than 20, symmetric distance, the top 5 non LOC / GPE entities, leaves, and are weighted unless stated otherwise. 

\begin{table} [h]
\centering
\caption{Positive Article Pairs for Evaluation Conditions (\%)}. 
\label{pos_eval}
\begin{tabular}{|l|r|r|r|}
\hline
 GR@.75 &  GR@.5 &  DR@.75 &  DR@.5 \\ \hline 
 25     &  40  & 8 & 21   \\ \hline
\end{tabular}
\vspace{-4mm} 
\end{table}

\subsection{Evaluation Conditions}

The two evaluation questions are highly subjective and the answers varied on an individual basis. For example, for question 2 only 15\% of article pairs were unanimously considered to be a good recommendation by all six participants. Therefore, in this work, we will evaluate our recommendation algorithms under the following four different conditions. In each condition, the {\it CNRec} data set is split into positive and negative samples in a different way to take the subjectivity issue into account.

\begin{enumerate}
 \item Good Recommendation at least XX\% ({\it GR@.50} and {\it GR@.75}): A pair of articles is considered as a good recommendation if the average rating score of Question 2 across all participants is 0.XX or above. 
 \item Diverse Recommendation at least XX\% ({\it DR@.50} and {\it DR@.75}): A pair of articles is considered as a good recommendation only if at least XX\% of participants think it is a good recommendation (Yes to Question 2) and 50\% of participants think they are not very similar (Question 1 rating is 1 or below). 
\end{enumerate}

From Table \ref{pos_eval} we can see that {\it GR@.50} will produce 40\% positive pairs. For {\it GR@.75} 25\% of article pairs are positive samples. Using {\it DR@.50} we have 21\% positive samples. Lastly for {\it DR@.75} we have 8\% as positive samples. 

In our experiments each algorithm's raw recommendation scores are normalized to be zero mean and unit variance. The scores are then judged against a universal threshold, 0 in this case, to make a binary recommendation decision. For each of the above conditions, the recommendation results are compared with the positive/negative labels to calculate precision and recall. The final recommendation performance is measured by $F_1$ score, i.e., 
$ F_1 = 2 * \frac{Precision * Recall}{Precision + Recall} $.

\subsection{Analysis on various hyperparameters}

In this section, we will study how the choices of some critical hyper-parameters in the proposed algorithm may affect the recommendation performance. 

\subsubsection{Optimal Maximal Length Expansion Radius L}

\begin{table}
\centering
\caption{Maximal Length Expansion Radius ($F_1$ scores in \%). 
(Penalty 0.98 was used in all tests. No LOC or GPE entities were used.)  }
\label{table-expand-rad}
\begin{tabular}{|lrrrr|}
\hline
Weighted / Hops             &  GR@.75 &  GR@.5 &  DR@.75 &  DR@.5 \\ \hline \hline
UnW / 2   &    64.57     &  70.23  & 23.16 & 38.26             \\
W / 2       &    64.87     &   71.62 & 23.79	& 39.75		\\  \hline
UnW / 1   &     64.45     &  72.07 & 23.51  & 40.37		\\  
W / 1      &    {\bf 65.09}     & {\bf 73.54} & {\bf 24.72}  & {\bf 42.49}          \\  \hline
\end{tabular}
\end{table}

From \ref{table-expand-rad} we can see that the one hop weighted subgraphs excel in all categories. The weighting is much more effective for the 1 hop graphs than the two hop graphs. We believe this is because our RW checks the immediate neighbors of two connecting nodes.  

\subsubsection{Effect of entity screening and context words}

As shown in Table \ref{table-comparion-entity-cw}, after entity pre-screening and removing all entities of the type of LOC and GPE, we observe large gains across four evaluation conditions for both unweighted and weighted distance calculation.  It is observed that while common nodes of LOC and GPE types may be useful in some situations, they contain too many neighboring nodes. This in turn, leads to many incorrect recommendations due to common entity overlapping. After removal of LOC and GPE there were 7.5 average nodes per article. Thus we skipped filtering top 10 entities in favor of top 8 and 5. Furthermore, we can see it is helpful to add a small number of context words for each article. Based on the results in  Table \ref{table-comparion-entity-cw}, we have decided to add up to 2 context words to each article since this yields the largest gain across various categories. 

\begin{table}
\centering
\caption{Entity pre-screening and context words performance comparison ($F_1$ scores in \%). 
(All 1 Hop; W: weighting;  NE: Number of Entities where NLG indicates No LOC or GPE. C: adding \# context words; Penalty 0.98 was used in all tests.)  }
\label{table-comparion-entity-cw}
\begin{tabular}{|lrrrr|}
\hline
W / NE / C        &  GR@.75 &  GR@.5 &  DR@.75 &  DR@.5 \\ \hline \hline
UnW / All (11 Avg) / 0     &    60.79      &          67.58              &         21.37 &       36.89             \\
UnW / NLG 8 / 0  &		64.45     &         72.07            & 			23.51			&     	40.37		\\  
UnW / NLG 5 / 0   &		72.53     &         78.46            & 			27.81			&     	44.98		\\
UnW / NLG 5 / 2 &		{\bf 75.89}     &         {\bf 83.76}            & 			{\bf 29.41}			&     	49.47		\\
UnW / NLG 5 / 4 &		74.30     &         84.25           & 			28.64			&     	{\bf 49.94}	\\  \hline 
W / All / 0         &    64.45      &         71.02               &         23.88              &       39.57          \\ 
W / NLG 8 / 0         &    65.09      &         73.54               &         24.72              &       42.49          \\ 
W / NLG 5  / 0  &		72.79 &                 80.02 &                28.42 &               46.84		\\
W / NLG 5 / 2    &		 {\bf 75.91} &                 {\bf 84.72} &                {\bf 29.41} &               49.76	\\
W / NLG 5 / 4    &		 74.29 &                 84.67 &                28.82 &               {\bf 50.66}		\\ \hline
\end{tabular}
\end{table}

\subsubsection{Effect of various edge weighting schemas}

\begin{figure}
	\centering
	\includegraphics[scale=0.225]{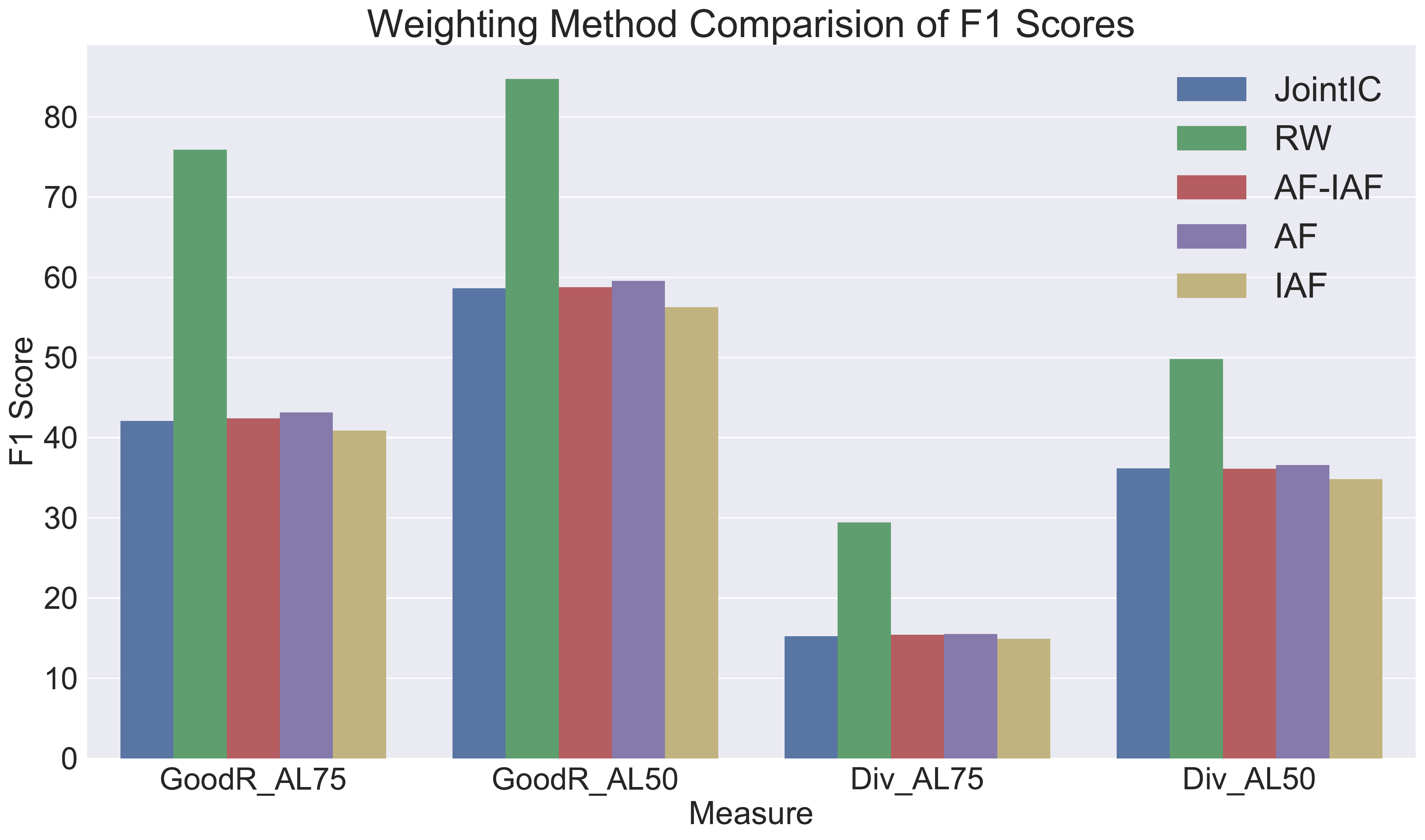}
	\caption{Edge weighting method comparison via Sym SED, entity pre-screening, 2 context words.}
	\label{Figure-comparison-weighingschema}
\end{figure}

Here we compare our proposed edge weighting method, RWS, in section \ref{subsection-RWC}, with four existing methods, i.e., AF, IAF and AF-IAF, JointIC, with SED. As shown in Figure \ref{Figure-comparison-weighingschema}, we can see that RWS produces much larger gains across all evaluation conditions over the three existing weighting methods. We believe this is because RWS weights each node based on the underlying article while the three traditional methods are purely based on term frequency over the entire knowledge graph. These methods likely impart less specific information on the active article. 

\subsubsection{Penalty for Disconnected Nodes}

\begin{table}
\centering
\begin{threeparttable}
\caption{SED Penalty Values: F$1$ Scores (\%)}
\label{Table-penalty}
\begin{tabular}{|lrrrr|}
\hline
{} &  GR@75 &  GR@50 & DR@75 &  DR@5  \\
\hline
NoP & 74.19 & 81.70 & 29.35 & 48.47 \\
P98 & {\bf 75.91} & {\bf 84.72} & {\bf 29.41}  &  {\bf 49.76} \\
P95 & 74.24 & 81.43 & 29.40 &   48.01 \\
P90 & 70.53 & 77.44 & 26.03 &   44.39 \\
\hline
\end{tabular}
\begin{tablenotes}
	\item{	
	\caption*{All tests use symmetric distance, filtered entities, 2 context words, and relational weighting. PXX indicates a penalty of 0.XX was used. }
	}
\end{tablenotes}
\end{threeparttable}
\vspace{-8mm} 
\end{table}

Our experiments from table \ref{Table-penalty} show by setting a lower penalty value, the overall accuracy of the algorithm is decreased. We believe this is because too many unrelated articles are recommended.

\subsection{Computation Time}

All graphs created are of expansion radius 1 which provided the best results for all algorithms. Using our filtered Freebase of 17.5 million nodes graphs were created in 50 milli-seconds (ms) on average. Weighting took 78.8 ms average per graph. The English only Freebase of 438 000 million nodes took 512 ms on average to create each graph. As EffGraph (EG) performed better using the unfiltered English only Freebase, it was used to create it's graphs. The average time for the 689 entity graph creation was 35 ms. Computing the all pairwise shortest distance to all other discovered entities (entities were skipped if not present in the graph) took 13.2 ms on average, although a large majority of entity pairs (99.26\%) were skipped. The 5400 evaluations took 29.4 ms on average for SED on the 17.5 million Freebase, 983 ms for the 438 000 SED graphs, 30.94 for KGD, and 0.00017 for EG. Computations used a Intel(R) Core(TM) i7-6500U CPU @ 2.50GHz, with 8 GB of RAM.

\begin{table*}
\caption{Recommendation Algorithm Performance.}
\label{Table-all-rec-comparison}
\begin{adjustbox}{max width=\textwidth}
\begin{tabular}{|ll|rr|rrrrr|}
\hline
      & &  \multicolumn{2}{|c|}{Conventional} &  \multicolumn{5}{|c|}{Graph Based} \\\hline
      & Algo &  TF-IDF & Doc2Vec &  EG &    KGD &  SED Avg P98 &  SED Row P98 &   SED Sym P98 \\\hline
DR@50 & F1 &    48.66 & {\bf 55.76} &     48.80 &  41.01 &        44.38 &        48.73 &              {\bf 49.76} \\
      & Precision &    63.29 & {\bf 91.43} &     63.81 &  58.74 &        55.59 &        64.60 &             {\bf 67.31}  \\
      & Recall &    39.52 & {\bf 40.11} &     {\bf 39.50} &  31.50 &        36.93 &        39.12 &           39.47 \\
\hline
DR@75 & F1 &    {\bf 29.28} & 27.19 &     {\bf 30.31} &  24.89 &        26.03 &        29.25 &          29.41 \\
      & Precision &    78.20 & {\bf 97.63} &     81.52 &  75.36 &        66.11 &        80.09 &               {\bf 82.70} \\
      & Recall &    {\bf 18.01} & 15.80 &     {\bf 18.61} &  14.91 &        16.20 &        17.89 &            17.89 \\
\hline
GR@50 & F1 &    85.56 & {\bf 85.89} &     83.81 &  73.68 &        77.43 &        83.21 &               {\bf 84.72} \\
      & Precision &    79.37 & {\bf 95.24} &     78.06 &  73.53 &        69.84 &        78.29 &             {\bf 80.95} \\
      & Recall &    {\bf 92.79} & 78.22 &     {\bf 90.48} &  73.84 &        86.88 &        88.78 &           88.88 \\
\hline
GR@75 & F1 &    {\bf 78.42} & 68.04 &     {\bf 77.04} &  66.82 &        70.53 &        75.27 &              75.91 \\
      & Precision &    91.80 & {\bf 98.98} &     90.63 &  85.58 &        79.72 &        89.68 &              {\bf 92.17} \\
      & Recall &    {\bf 68.45} & 51.84 &     {\bf 66.99} &  54.81 &        63.24 &        64.85 &            64.53 \\
\hline
\end{tabular}
\end{adjustbox}
\end{table*}

\begin{table}
%\scalebox{0.75} 
\caption{Recommendation Algorithm Ensembles Performance.}
%\text{Recommendation Algorithm Ensembles Performance.}
\begin{adjustbox}{max width=\textwidth}
\begin{tabular}{|ll|rrrrrrr|}
\hline
      & Algo &  EG Doc &  EG Doc TF &  EG TF &  SED Doc &  SED Doc TF &  SED TF &  TF Doc \\ \hline
DR@50 & F1 &      56.48 &           55.73 &     47.93 &      {\bf 57.19} &           56.17 &     49.03 &     56.49 \\
      & Precision &      {\bf 88.99} &           78.41 &     61.71 &      88.29 &           79.81 &     63.81 &     81.03 \\
      & Recall &      41.37 &           43.23 &     39.18 &      42.29 &           43.33 &     39.80 &     {\bf 43.36} \\ \hline
DR@75 & F1 &      28.37 &           {\bf 31.48} &     30.04 &      29.25 &           31.47 &     30.59 &     30.86 \\
      & Precision &      96.92 &           93.13 &     79.15 &      {\bf 97.39} &           94.31 &     81.75 &     93.60 \\
      & Recall &      16.62 &           {\bf 18.94} &     18.53 &      17.21 &           18.89 &     18.81 &     18.48 \\ \hline
GR@50 & F1 &      87.42 &           89.45 &     85.40 &      88.39 &           {\bf 89.62} &     86.02 &     89.49 \\
      & Precision &      {\bf 93.93} &           88.05 &     78.62 &      93.46 &           88.89 &     79.83 &     89.40 \\
      & Recall &      81.76 &           90.89 &     {\bf 93.45} &      83.84 &           90.37 &     93.24 &     89.57 \\ \hline
GR@75 & F1 &      70.50 &           77.19 &     79.42 &      71.87 &           76.88 &     {\bf 79.44} &     75.86 \\
      & Precision &      98.76 &           97.22 &     92.09 &      {\bf 98.76} &           97.73 &     93.05 &     97.29 \\
      & Recall &      54.82 &           64.00 &     {\bf 69.81} &      56.49 &           63.36 &     69.30 &     62.16 \\
\hline
\end{tabular} 
\end{adjustbox} 
\end{table}

\subsection{Baseline Parameters}

For experiments utilizing TF-IDF all stop words are removed from the corpus. We then filter out any words in over 80\% of articles. The Doc2Vec model is the distributed memory (PV-DM) model. Past research has shown that PV-DM outperforms the the distributed bag of words (PV-DBOW) model when the corpus is small. As our data set is 300 articles this is a reasonable choice. Our embedding size is set to 300. 

As SED computes the distance between two articles TF-IDF and doc2Vec were converted from similarity to distance measures. Since both similarity computations were done with cosine similarity they output values between $-1$ and $1$. To bound the results between 0 and 1 any values that were less than 0, indicating that the vectors are diametrically opposed, were set to 0. Then to convert from similarity to distance we subtracted the similarity value from 1. 

\subsection{Main Results}

In Table \ref{Table-all-rec-comparison}, we show the performance of two baseline recommendation algorithms {\it TF-IDF} and {\it Doc2vec}, in terms of $F_1$ scores for all four evaluation conditions. From there, we see that {\it TF-IDF} outperforms {\it Doc2vec} by a large margin in {\it GR@.75 } and {\it DR@.75} while {\it Doc2vec} displays higher accuracies in the at-least-50\% tests.

\begin{figure}
	\centering
	\includegraphics[scale=0.40]{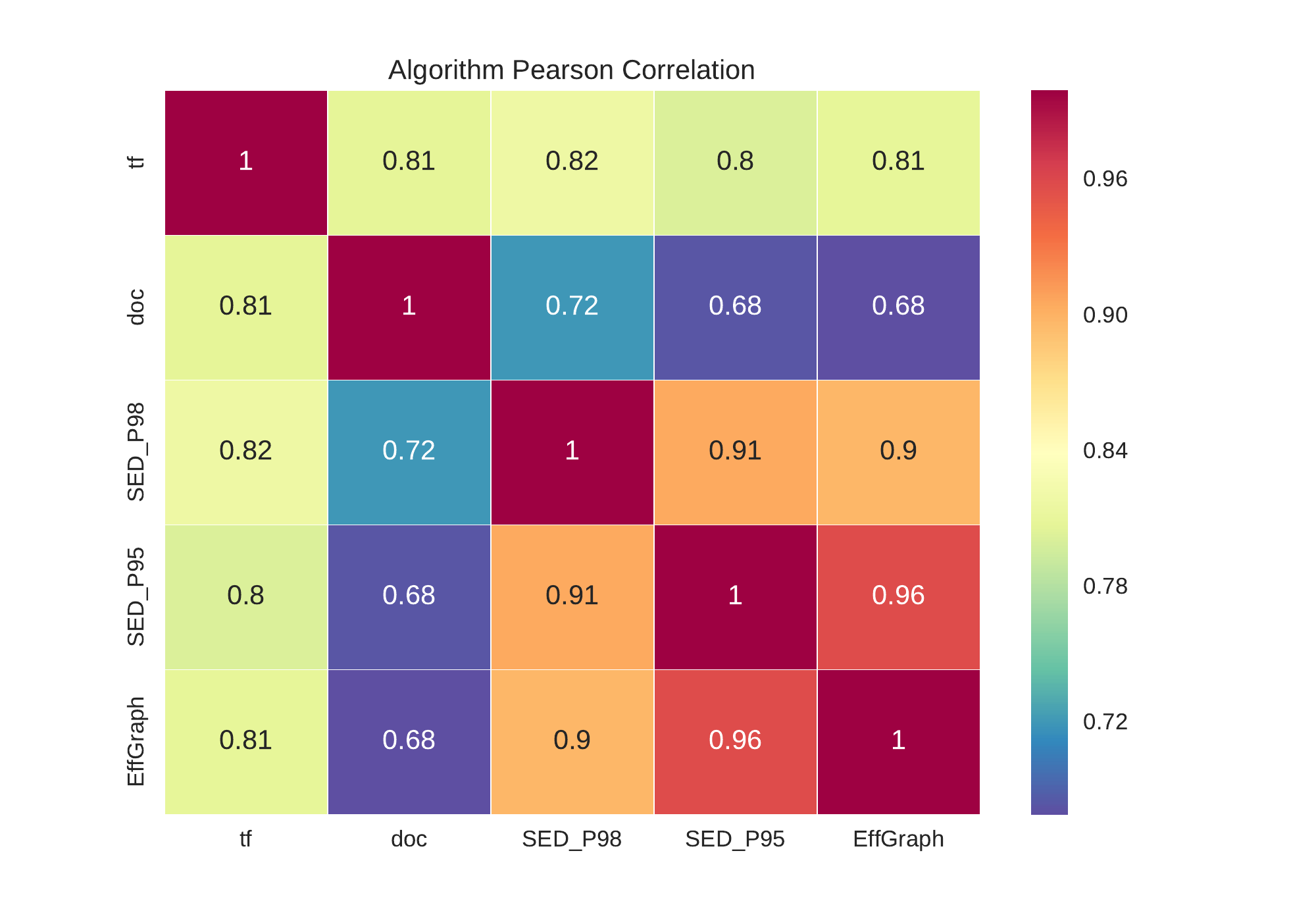}
	\caption{Pearson correlation of recommendation scores from various algorithms. }
	\label{Figure-correlation-diff-algorithm}
	\vspace{-4mm}%Put here to reduce too much white space after your table 
\end{figure}

Next, we have evaluated several configurations of the proposed shortest entity distance (SED) method for the four evaluation conditions. From the results in Table \ref{Table-all-rec-comparison}, it is clear that the \cite{KGD} bipartite matching and all-average distance calculation performs the worst in all cases. This may be due to the fact that the distance measures from some critical entities are over-averaged or ignored due to prior matchings by other less importance entities. The symmetric distance computation yields the best performance across all categories. We believe this is because the symmetry prevents one article with many entities from encapsulating the other. 

By itself {\it SED} symmetric performs much better than {\it Doc2Vec} in {\it GR@.75} and  {\it DR@.75} categories. It gives higher results versus {\it EG} in both {\it @ 0.5} categories. It  also provides better performance in the {\it DR@.75} case over {\it TF-IDF} and {\it Doc2Vec} although lower than {\it EffGraph}. In the {\it DR@.50} category, the SED symmetric variant outperforms {\it TF-IDF} and {\it EG} but is far behind {\it doc2vec}. From these observations, it may be concluded that SED leverages additional information from both the context and entity nodes to make more refined diverse recommendations than the baseline algorithms, but less refined versus {\it EG}. 

As shown in Figure \ref{Figure-correlation-diff-algorithm} we can see that the SED correlates less with {\it Doc2vec} and {\it TF-IDF} than they do with each other. The results in Table \ref{Table-all-rec-comparison} show that a consolidation of SED, TF-IDF, and doc2vec provides the highest performance in {\it GR@.75} producing a gain of $1.02\%$ over the blending of {\it TF-IDF} and {\it doc2vec}. A combination of SED and {\it doc2vec} provides the highest overall performance in {\it GR@.5} producing a gain of $0.13\%$  over the second best of {\it TF-IDF} and {\it doc2vec} . A synthesis of SED and {\it TF-IDF} provides the highest performance in {\it DR@.75} producing a gain of $0.62\%$ over the second best of {\it TF-IDF} and {\it doc2vec}. A combination of SED and {\it doc2vec} give the best performance in {\it DR@.5}.  This accounts for a gain of $0.70\%$ over the second best of {\it TF-IDF} and {\it doc2vec}. 
Our hypothesis is that the smaller correlation between SED and the baseline algorithms accounts for the gains as SED would recommend articles the baseline algorithms would not. 

Finally, we evaluate the correlation between the algorithm's recommendation scores and the average annotation scores of Question 2 among the six human annotators. In Table \ref{Table-correlation-humanscore}, we first list the correlation coefficients (both Pearson and Spearman) of the baseline algorithms: {\it TF-IDF} and {\it Doc2vec}.  As we can see, the correlation coefficients of SED alone is much lower than both baseline algorithms.  Only 5.3\% of distance values returned a 1 when using SED. While many other values returned close to 1, these values do no correlate to the human similarity scores that were exactly 1. When using EG 63.1\% of distance values were 1. We believe this is why SED has higher precision VS EG in all categories and produces a higher Pearson correlation versus both other traversal algorithms, but lower Spearman VS EG.

\begin{table}
\centering
\caption{Correlation coefficients between average human annotation scores of Question 2 and recommendation scores from various algorithms.}
\label{Table-correlation-humanscore}
\begin{tabular}{|lrr|}
\hline
 &    Pearson &   Spearman \\
\hline
Doc2Vec        &  78.832 &  81.155 \\
TF-IDF         &  {\bf 87.709} &  {\bf 81.469} \\ \hline \hline
KGD & 70.093 & 61.413 \\
EG & 78.975 & {\bf 78.870} \\
SED  &  {\bf 79.273} &  72.730 \\  \hline \hline
TF + Doc    &  88.310 &  83.456 \\ 
SED + TF        &  88.562 &  81.899 \\
SED + Doc   &  84.271 &  82.642 \\
SED + Doc + IF &  {\bf 88.957}  &  {\bf 83.852} \\ \hline 
\end{tabular}
\vspace{-4mm}
\end{table}

\section{Conclusions}

The proposed shortest entity distance (SED) algorithm performs the best under almost all examined conditions in the cold-start content-based news recommendation task when it is configured with maximal length 1 expansion, relation weighting scheme, entity pre-screening, leaf nodes, and up to 2 context words. Moreover, the above SED algorithm yields significant gains in these tasks when combined with some conventional recommendation algorithms. This has shown that the proposed SED algorithm has effectively leveraged useful information from knowledge graphs for content-based recommendation, which is fairly complementary with most existing algorithms. SED also provides the highest Pearson correlation of graph traversal algorithms. 

Lastly, we have created a new data set for content-based news recommendation from this work. We hope that this data set will become a standard benchmark in the content-based news recommendation field as it provides a variety of human opinions and a diverse set of answers allowing researchers to evaluate their algorithms across many metrics. 

\newpage
\bibliographystyle{ACM-Reference-Format}
\balance 
\bibliography{CNRec_WWW}{}


\begin{thebibliography}{35}

%%% the Web address.

\ifx \showCODEN    \undefined \def \showCODEN     #1{\unskip}     \fi
\ifx \showDOI      \undefined \def \showDOI       #1{#1}\fi
\ifx \showISBNx    \undefined \def \showISBNx     #1{\unskip}     \fi
\ifx \showISBNxiii \undefined \def \showISBNxiii  #1{\unskip}     \fi
\ifx \showISSN     \undefined \def \showISSN      #1{\unskip}     \fi
\ifx \showLCCN     \undefined \def \showLCCN      #1{\unskip}     \fi
\ifx \shownote     \undefined \def \shownote      #1{#1}          \fi
\ifx \showarticletitle \undefined \def \showarticletitle #1{#1}   \fi
\ifx \showURL      \undefined \def \showURL       {\relax}        \fi
% invisible to TeX
\providecommand\bibfield[2]{#2}
\providecommand\bibinfo[2]{#2}
\providecommand\natexlab[1]{#1}
\providecommand\showeprint[2][]{arXiv:#2}

\bibitem[\protect\citeauthoryear{Auer, Bizer, Kobilarov, Lehmann, Cyganiak, and
  Ives}{Auer et~al\mbox{.}}{2007}]%
        {Auer:2007:DNW:1785162.1785216}
\bibfield{author}{\bibinfo{person}{S\"{o}ren Auer}, \bibinfo{person}{Christian
  Bizer}, \bibinfo{person}{Georgi Kobilarov}, \bibinfo{person}{Jens Lehmann},
  \bibinfo{person}{Richard Cyganiak}, {and} \bibinfo{person}{Zachary Ives}.}
  \bibinfo{year}{2007}\natexlab{}.
\newblock \showarticletitle{DBpedia: A Nucleus for a Web of Open Data}. In
  \bibinfo{booktitle}{\emph{Proceedings of the 6th International The Semantic
  Web and 2Nd Asian Conference on Asian Semantic Web Conference}}
  \emph{(\bibinfo{series}{ISWC'07/ASWC'07})}.
  \bibinfo{publisher}{Springer-Verlag}, \bibinfo{address}{Berlin, Heidelberg},
  \bibinfo{pages}{722--735}.
\newblock
\showISBNx{3-540-76297-3, 978-3-540-76297-3}
\urldef\tempurl%
\url{http://dl.acm.org/citation.cfm?id=1785162.1785216}
\showURL{%
\tempurl}


\bibitem[\protect\citeauthoryear{Bollacker, Evans, Paritosh, Sturge, and
  Taylor}{Bollacker et~al\mbox{.}}{2008}]%
        {freebase:datadumps}
\bibfield{author}{\bibinfo{person}{Kurt Bollacker}, \bibinfo{person}{Colin
  Evans}, \bibinfo{person}{Praveen Paritosh}, \bibinfo{person}{Tim Sturge},
  {and} \bibinfo{person}{Jamie Taylor}.} \bibinfo{year}{2008}\natexlab{}.
\newblock \showarticletitle{Freebase: A Collaboratively Created Graph Database
  for Structuring Human Knowledge}. In \bibinfo{booktitle}{\emph{Proceedings of
  the 2008 ACM SIGMOD International Conference on Management of Data}}
  \emph{(\bibinfo{series}{SIGMOD '08})}. \bibinfo{publisher}{ACM},
  \bibinfo{address}{New York, NY, USA}, \bibinfo{pages}{1247--1250}.
\newblock
\showISBNx{978-1-60558-102-6}
\urldef\tempurl%
\url{https://doi.org/10.1145/1376616.1376746}
\showDOI{\tempurl}


\bibitem[\protect\citeauthoryear{Das, Datar, Garg, and Rajaram}{Das
  et~al\mbox{.}}{2007}]%
        {Das:2007:GNP:1242572.1242610}
\bibfield{author}{\bibinfo{person}{Abhinandan~S. Das}, \bibinfo{person}{Mayur
  Datar}, \bibinfo{person}{Ashutosh Garg}, {and} \bibinfo{person}{Shyam
  Rajaram}.} \bibinfo{year}{2007}\natexlab{}.
\newblock \showarticletitle{Google News Personalization: Scalable Online
  Collaborative Filtering}. In \bibinfo{booktitle}{\emph{Proceedings of the
  16th International Conference on World Wide Web}} \emph{(\bibinfo{series}{WWW
  '07})}. \bibinfo{publisher}{ACM}, \bibinfo{address}{New York, NY, USA},
  \bibinfo{pages}{271--280}.
\newblock
\showISBNx{978-1-59593-654-7}
\urldef\tempurl%
\url{https://doi.org/10.1145/1242572.1242610}
\showDOI{\tempurl}


\bibitem[\protect\citeauthoryear{E~Ali and Lawless}{E~Ali and Lawless}{2017}]%
        {Esraa}
\bibfield{author}{\bibinfo{person}{A~Caputo E~Ali} {and} \bibinfo{person}{S
  Lawless}.} \bibinfo{year}{2017}\natexlab{}.
\newblock \showarticletitle{Entity Attribute Ranking Using Learning to Rank}.
\newblock
  \bibinfo{howpublished}{\url{http://ceur-ws.org/Vol-1883/paper_10.pdf}}.
\newblock \bibinfo{journal}{\emph{CEUR}}  \bibinfo{volume}{1883}
  (\bibinfo{year}{2017}).
\newblock


\bibitem[\protect\citeauthoryear{Garcin, Dimitrakakis, and Faltings}{Garcin
  et~al\mbox{.}}{2013}]%
        {Garcin:2013:PNR:2507157.2507166}
\bibfield{author}{\bibinfo{person}{Florent Garcin}, \bibinfo{person}{Christos
  Dimitrakakis}, {and} \bibinfo{person}{Boi Faltings}.}
  \bibinfo{year}{2013}\natexlab{}.
\newblock \showarticletitle{Personalized News Recommendation with Context
  Trees}. In \bibinfo{booktitle}{\emph{Proceedings of the 7th ACM Conference on
  Recommender Systems}} \emph{(\bibinfo{series}{RecSys '13})}.
  \bibinfo{publisher}{ACM}, \bibinfo{address}{New York, NY, USA},
  \bibinfo{pages}{105--112}.
\newblock
\showISBNx{978-1-4503-2409-0}
\urldef\tempurl%
\url{https://doi.org/10.1145/2507157.2507166}
\showDOI{\tempurl}


\bibitem[\protect\citeauthoryear{Hangal, MacLean, Lam, and Heer}{Hangal
  et~al\mbox{.}}{2010}]%
        {SocialInfluence}
\bibfield{author}{\bibinfo{person}{Sudheendra Hangal},
  \bibinfo{person}{Diana~L. MacLean}, \bibinfo{person}{Monica~S. Lam}, {and}
  \bibinfo{person}{Jeffrey Heer}.} \bibinfo{year}{2010}\natexlab{}.
\newblock \showarticletitle{All Friends are Not Equal: Using Weights in Social
  Graphs to Improve Search}.
  \bibinfo{howpublished}{\url{http://vis.stanford.edu/files/2010-WeightsSocialGraphs-SNAKDD.pdf}}.
\newblock


\bibitem[\protect\citeauthoryear{Heitmann and Hayes}{Heitmann and
  Hayes}{2016}]%
        {SemStim}
\bibfield{author}{\bibinfo{person}{B. Heitmann} {and} \bibinfo{person}{C.
  Hayes}.} \bibinfo{year}{2016}\natexlab{}.
\newblock \showarticletitle{SemStim: Exploiting Knowledge Graphs for
  Cross-Domain Recommendation}. In \bibinfo{booktitle}{\emph{2016 IEEE 16th
  International Conference on Data Mining Workshops (ICDMW)}}.
  \bibinfo{pages}{999--1006}.
\newblock
\urldef\tempurl%
\url{https://doi.org/10.1109/ICDMW.2016.0145}
\showDOI{\tempurl}


\bibitem[\protect\citeauthoryear{Hulpus, Hayes, Karnstedt, and Greene}{Hulpus
  et~al\mbox{.}}{2013}]%
        {UnsupGraphTopic}
\bibfield{author}{\bibinfo{person}{Ioana Hulpus}, \bibinfo{person}{Conor
  Hayes}, \bibinfo{person}{Marcel Karnstedt}, {and} \bibinfo{person}{Derek
  Greene}.} \bibinfo{year}{2013}\natexlab{}.
\newblock \showarticletitle{Unsupervised Graph-based Topic Labelling Using
  Dbpedia}. In \bibinfo{booktitle}{\emph{Proceedings of the Sixth ACM
  International Conference on Web Search and Data Mining}}
  \emph{(\bibinfo{series}{WSDM '13})}. \bibinfo{publisher}{ACM},
  \bibinfo{address}{New York, NY, USA}, \bibinfo{pages}{465--474}.
\newblock
\showISBNx{978-1-4503-1869-3}
\urldef\tempurl%
\url{https://doi.org/10.1145/2433396.2433454}
\showDOI{\tempurl}


\bibitem[\protect\citeauthoryear{Ji and Nothman}{Ji and Nothman}{2016}]%
        {tac2016overview}
\bibfield{author}{\bibinfo{person}{Heng Ji} {and} \bibinfo{person}{Joel
  Nothman}.} \bibinfo{year}{2016}\natexlab{}.
\newblock \showarticletitle{Overview of TAC-KBP 2016 Tri-lingual EDL and Its
  Impact on End-to-End Cold-Start KBP}. In
  \bibinfo{booktitle}{\emph{Proceedings of 2016 Text Analysis Conference
  (TAC)}}.
\newblock
\urldef\tempurl%
\url{https://tac.nist.gov/publications/2016/papers.html}
\showURL{%
\tempurl}


\bibitem[\protect\citeauthoryear{Kille, Lommatzsch, Hopfgartner, Larson, and
  Brodt}{Kille et~al\mbox{.}}{2017}]%
        {LiveTesting}
\bibfield{author}{\bibinfo{person}{Benjamin Kille}, \bibinfo{person}{Andreas
  Lommatzsch}, \bibinfo{person}{Frank Hopfgartner}, \bibinfo{person}{Martha
  Larson}, {and} \bibinfo{person}{Torben Brodt}.}
  \bibinfo{year}{2017}\natexlab{}.
\newblock \bibinfo{title}{CLEF 2017 NewsREEL Overview: Offline and Online
  Evaluation of Stream-based News Recommender Systems}.
\newblock
\newblock
\urldef\tempurl%
\url{http://eprints.gla.ac.uk/143586/}
\showURL{%
\tempurl}


\bibitem[\protect\citeauthoryear{Lam}{Lam}{2013}]%
        {DbLam}
\bibfield{author}{\bibinfo{person}{Samantha Lam}.}
  \bibinfo{year}{2013}\natexlab{}.
\newblock \showarticletitle{Using the Structure of DBpedia for Exploratory
  Search}.
  \bibinfo{howpublished}{\url{http://chbrown.github.io/kdd-2013-usb/workshops/MDS/doc/mds2013_submission_2.pdf}}.
\newblock


\bibitem[\protect\citeauthoryear{Lau and Baldwin}{Lau and Baldwin}{2016}]%
        {DBLP:journals/corr/LauB16}
\bibfield{author}{\bibinfo{person}{Jey~Han Lau} {and} \bibinfo{person}{Timothy
  Baldwin}.} \bibinfo{year}{2016}\natexlab{}.
\newblock \showarticletitle{An Empirical Evaluation of doc2vec with Practical
  Insights into Document Embedding Generation}.
\newblock \bibinfo{journal}{\emph{CoRR}}  \bibinfo{volume}{abs/1607.05368}
  (\bibinfo{year}{2016}).
\newblock
\urldef\tempurl%
\url{http://arxiv.org/abs/1607.05368}
\showURL{%
\tempurl}


\bibitem[\protect\citeauthoryear{Le and Mikolov}{Le and Mikolov}{2014}]%
        {DBLP:journals/corr/LeM14}
\bibfield{author}{\bibinfo{person}{Quoc~V. Le} {and} \bibinfo{person}{Tomas
  Mikolov}.} \bibinfo{year}{2014}\natexlab{}.
\newblock \showarticletitle{Distributed Representations of Sentences and
  Documents}.
\newblock \bibinfo{journal}{\emph{CoRR}}  \bibinfo{volume}{abs/1405.4053}
  (\bibinfo{year}{2014}).
\newblock
\urldef\tempurl%
\url{http://arxiv.org/abs/1405.4053}
\showURL{%
\tempurl}


\bibitem[\protect\citeauthoryear{Leal, Rodrigues, and Queirós}{Leal
  et~al\mbox{.}}{2012}]%
        {DbLeal}
\bibfield{author}{\bibinfo{person}{José~Paulo Leal}, \bibinfo{person}{Vânia
  Rodrigues}, {and} \bibinfo{person}{Ricardo Queirós}.}
  \bibinfo{year}{2012}\natexlab{}.
\newblock \showarticletitle{Computing Semantic Relatedness using DBPedia.}. In
  \bibinfo{booktitle}{\emph{SLATE}} \emph{(\bibinfo{series}{OASICS})},
  \bibfield{editor}{\bibinfo{person}{Alberto Simões}, \bibinfo{person}{Ricardo
  Queirós}, {and} \bibinfo{person}{Daniela~Carneiro da~Cruz}} (Eds.),
  Vol.~\bibinfo{volume}{21}. \bibinfo{publisher}{Schloss Dagstuhl -
  Leibniz-Zentrum fuer Informatik}, \bibinfo{pages}{133--147}.
\newblock
\showISBNx{978-3-939897-40-8}
\urldef\tempurl%
\url{http://dblp.uni-trier.de/db/conf/slate/slate2012.html#LealRQ12}
\showURL{%
\tempurl}


\bibitem[\protect\citeauthoryear{Lee and Welsh}{Lee and Welsh}{2005}]%
        {LG50}
\bibfield{author}{\bibinfo{person}{Michael~D. Lee} {and}
  \bibinfo{person}{Matthew Welsh}.} \bibinfo{year}{2005}\natexlab{}.
\newblock \showarticletitle{An empirical evaluation of models of text document
  similarity}.
  \bibinfo{howpublished}{\url{http://www.socsci.uci.edu/~mdlee/lee_pincombe_welsh_document.PDF}}.
  In \bibinfo{booktitle}{\emph{In CogSci2005}}. \bibinfo{publisher}{Erlbaum},
  \bibinfo{pages}{1254--1259}.
\newblock


\bibitem[\protect\citeauthoryear{Li, Wang, Zhu, and Li}{Li
  et~al\mbox{.}}{2011}]%
        {Li:2011:PNR:2336266.2336268}
\bibfield{author}{\bibinfo{person}{Lei Li}, \bibinfo{person}{Ding-Ding Wang},
  \bibinfo{person}{Shun-Zhi Zhu}, {and} \bibinfo{person}{Tao Li}.}
  \bibinfo{year}{2011}\natexlab{}.
\newblock \showarticletitle{Personalized News Recommendation: A Review and an
  Experimental Investigation}.
\newblock \bibinfo{journal}{\emph{J. Comput. Sci. Technol.}}
  \bibinfo{volume}{26}, \bibinfo{number}{5} (\bibinfo{date}{Sept.}
  \bibinfo{year}{2011}), \bibinfo{pages}{754--766}.
\newblock
\showISSN{1000-9000}
\urldef\tempurl%
\url{https://doi.org/10.1007/s11390-011-0175-2}
\showDOI{\tempurl}


\bibitem[\protect\citeauthoryear{Liu, Lin, Zhang, Wei, and Jiang}{Liu
  et~al\mbox{.}}{[n. d.]}]%
        {iflytek:edl:2016}
\bibfield{author}{\bibinfo{person}{Dan Liu}, \bibinfo{person}{Wei Lin},
  \bibinfo{person}{Shiliang Zhang}, \bibinfo{person}{Si Wei}, {and}
  \bibinfo{person}{Hui Jiang}.} \bibinfo{year}{[n. d.]}\natexlab{}.
\newblock \showarticletitle{Neural Networks Models for Entity Discovery and
  Linking}. In \bibinfo{booktitle}{\emph{arXiv:1611.03558}}.
\newblock
\urldef\tempurl%
\url{https://arxiv.org/abs/1611.03558}
\showURL{%
\tempurl}


\bibitem[\protect\citeauthoryear{Miller}{Miller}{1995}]%
        {Miller:1995:WLD:219717.219748}
\bibfield{author}{\bibinfo{person}{George~A. Miller}.}
  \bibinfo{year}{1995}\natexlab{}.
\newblock \showarticletitle{WordNet: A Lexical Database for English}.
\newblock \bibinfo{journal}{\emph{Commun. ACM}} \bibinfo{volume}{38},
  \bibinfo{number}{11} (\bibinfo{date}{Nov.} \bibinfo{year}{1995}),
  \bibinfo{pages}{39--41}.
\newblock
\showISSN{0001-0782}
\urldef\tempurl%
\url{https://doi.org/10.1145/219717.219748}
\showDOI{\tempurl}


\bibitem[\protect\citeauthoryear{Mingbin, Wei, Watcharawittayakul, Kang, and
  Jiang}{Mingbin et~al\mbox{.}}{2016}]%
        {yorkU:edl:2016}
\bibfield{author}{\bibinfo{person}{Mingbin}, \bibinfo{person}{Feng Wei},
  \bibinfo{person}{Sedtawut Watcharawittayakul}, \bibinfo{person}{Yuchen Kang},
  {and} \bibinfo{person}{Hui Jiang}.} \bibinfo{year}{2016}\natexlab{}.
\newblock \showarticletitle{The YorkNRM Systems for Trilingual EDL Tasks at TAC
  KBP 2016}. In \bibinfo{booktitle}{\emph{Proceedings of 2016 Text Analysis
  Conference (TAC)}}.
\newblock
\urldef\tempurl%
\url{https://tac.nist.gov/publications/2016/papers.html}
\showURL{%
\tempurl}


\bibitem[\protect\citeauthoryear{Nunes, Kawase, Fetahu, Dietze, Casanova, and
  Maynard}{Nunes et~al\mbox{.}}{2013}]%
        {SCS}
\bibfield{author}{\bibinfo{person}{Bernardo~Pereira Nunes},
  \bibinfo{person}{Ricardo Kawase}, \bibinfo{person}{Besnik Fetahu},
  \bibinfo{person}{Stefan Dietze}, \bibinfo{person}{Marco~A. Casanova}, {and}
  \bibinfo{person}{Diana Maynard}.} \bibinfo{year}{2013}\natexlab{}.
\newblock \showarticletitle{Interlinking Documents based on Semantic Graphs}.
\newblock \bibinfo{journal}{\emph{Procedia Computer Science}}
  \bibinfo{volume}{22}, \bibinfo{number}{Supplement C} (\bibinfo{year}{2013}),
  \bibinfo{pages}{231 -- 240}.
\newblock
\showISSN{1877-0509}
\urldef\tempurl%
\url{https://doi.org/10.1016/j.procs.2013.09.099}
\showDOI{\tempurl}
\newblock
\shownote{17th International Conference in Knowledge Based and Intelligent
  Information and Engineering Systems - KES2013.}


\bibitem[\protect\citeauthoryear{Ozgobek, Gulla, and Erdur}{Ozgobek
  et~al\mbox{.}}{2014}]%
        {webist14}
\bibfield{author}{\bibinfo{person}{Ozlem Ozgobek}, \bibinfo{person}{Jon~Atle
  Gulla}, {and} \bibinfo{person}{R.~Cenk Erdur}.}
  \bibinfo{year}{2014}\natexlab{}.
\newblock \showarticletitle{A Survey on Challenges and Methods in News
  Recommendation}. In \bibinfo{booktitle}{\emph{Proceedings of the 10th
  International Conference on Web Information Systems and Technologies - Volume
  2: WEBIST,}}. INSTICC, \bibinfo{publisher}{SciTePress},
  \bibinfo{pages}{278--285}.
\newblock
\showISBNx{978-989-758-024-6}
\urldef\tempurl%
\url{https://doi.org/10.5220/0004844202780285}
\showDOI{\tempurl}


\bibitem[\protect\citeauthoryear{Palma, Vidal, Haag, Raschid, and Thor}{Palma
  et~al\mbox{.}}{2013}]%
        {AnnSim}
\bibfield{author}{\bibinfo{person}{Guillermo Palma},
  \bibinfo{person}{Maria-Esther Vidal}, \bibinfo{person}{Eric Haag},
  \bibinfo{person}{Louiqa Raschid}, {and} \bibinfo{person}{Andreas Thor}.}
  \bibinfo{year}{2013}\natexlab{}.
\newblock \bibinfo{title}{Measuring Relatedness Between Scientific Entities in
  Annotation Datasets}.
\newblock \bibinfo{howpublished}{\url{https://bit.ly/2HAyeVA}}.
\newblock


\bibitem[\protect\citeauthoryear{Passant}{Passant}{2010a}]%
        {DbMusic}
\bibfield{author}{\bibinfo{person}{Alexandre Passant}.}
  \bibinfo{year}{2010}\natexlab{a}.
\newblock \showarticletitle{Dbrec: Music Recommendations Using DBpedia}. In
  \bibinfo{booktitle}{\emph{Proceedings of the 9th International Semantic Web
  Conference on The Semantic Web - Volume Part II}}
  \emph{(\bibinfo{series}{ISWC'10})}. \bibinfo{publisher}{Springer-Verlag},
  \bibinfo{address}{Berlin, Heidelberg}, \bibinfo{pages}{209--224}.
\newblock
\showISBNx{3-642-17748-4, 978-3-642-17748-4}
\urldef\tempurl%
\url{http://dl.acm.org/citation.cfm?id=1940334.1940349}
\showURL{%
\tempurl}


\bibitem[\protect\citeauthoryear{Passant}{Passant}{2010b}]%
        {LDSD}
\bibfield{author}{\bibinfo{person}{Alexandre Passant}.}
  \bibinfo{year}{2010}\natexlab{b}.
\newblock \bibinfo{title}{Measuring Semantic Distance on Linking Data and Using
  it for Resources Recommendations}.
\newblock
  \bibinfo{howpublished}{\url{https://www.aaai.org/ocs/index.php/SSS/SSS10/paper/view/1147}}.
\newblock


\bibitem[\protect\citeauthoryear{Paul, Rettinger, Mogadala, Knoblock, and
  Szekely}{Paul et~al\mbox{.}}{2016}]%
        {EfficientGraph}
\bibfield{author}{\bibinfo{person}{Christian Paul}, \bibinfo{person}{Achim
  Rettinger}, \bibinfo{person}{Aditya Mogadala}, \bibinfo{person}{Craig~A.
  Knoblock}, {and} \bibinfo{person}{Pedro Szekely}.}
  \bibinfo{year}{2016}\natexlab{}.
\newblock \showarticletitle{Efficient Graph-Based Document Similarity}. In
  \bibinfo{booktitle}{\emph{Proceedings of the 13th International Conference on
  The Semantic Web. Latest Advances and New Domains - Volume 9678}}.
  \bibinfo{publisher}{Springer-Verlag New York, Inc.}, \bibinfo{address}{New
  York, NY, USA}, \bibinfo{pages}{334--349}.
\newblock
\showISBNx{978-3-319-34128-6}
\urldef\tempurl%
\url{https://doi.org/10.1007/978-3-319-34129-3_21}
\showDOI{\tempurl}


\bibitem[\protect\citeauthoryear{Ramos}{Ramos}{2003}]%
        {Ramos2003UsingTT}
\bibfield{author}{\bibinfo{person}{Juan Ramos}.}
  \bibinfo{year}{2003}\natexlab{}.
\newblock \showarticletitle{Using TF-IDF to Determine Word Relevance in
  Document Queries}. \bibinfo{howpublished}{\url{https://bit.ly/2HRlE7J}}.
\newblock


\bibitem[\protect\citeauthoryear{Sanfeliu and Fu}{Sanfeliu and Fu}{1983}]%
        {GED}
\bibfield{author}{\bibinfo{person}{A. Sanfeliu} {and} \bibinfo{person}{K.~S.
  Fu}.} \bibinfo{year}{1983}\natexlab{}.
\newblock \showarticletitle{A distance measure between attributed relational
  graphs for pattern recognition}.
\newblock \bibinfo{journal}{\emph{IEEE Transactions on Systems, Man, and
  Cybernetics}} \bibinfo{volume}{SMC-13}, \bibinfo{number}{3}
  (\bibinfo{date}{May} \bibinfo{year}{1983}), \bibinfo{pages}{353--362}.
\newblock
\showISSN{0018-9472}
\urldef\tempurl%
\url{https://doi.org/10.1109/TSMC.1983.6313167}
\showDOI{\tempurl}


\bibitem[\protect\citeauthoryear{Schuhmacher and Ponzetto}{Schuhmacher and
  Ponzetto}{2014}]%
        {KGD}
\bibfield{author}{\bibinfo{person}{Michael Schuhmacher} {and}
  \bibinfo{person}{Simone~Paolo Ponzetto}.} \bibinfo{year}{2014}\natexlab{}.
\newblock \showarticletitle{Knowledge-based Graph Document Modeling}. In
  \bibinfo{booktitle}{\emph{Proceedings of the 7th ACM International Conference
  on Web Search and Data Mining}} \emph{(\bibinfo{series}{WSDM '14})}.
  \bibinfo{publisher}{ACM}, \bibinfo{address}{New York, NY, USA},
  \bibinfo{pages}{543--552}.
\newblock
\showISBNx{978-1-4503-2351-2}
\urldef\tempurl%
\url{https://doi.org/10.1145/2556195.2556250}
\showDOI{\tempurl}


\bibitem[\protect\citeauthoryear{Shi, Kong, Huang, Yu, and Wu}{Shi
  et~al\mbox{.}}{2013}]%
        {HeteSim}
\bibfield{author}{\bibinfo{person}{Chuan Shi}, \bibinfo{person}{Xiangnan Kong},
  \bibinfo{person}{Yue Huang}, \bibinfo{person}{Philip~S. Yu}, {and}
  \bibinfo{person}{Bin Wu}.} \bibinfo{year}{2013}\natexlab{}.
\newblock \showarticletitle{HeteSim: A General Framework for Relevance Measure
  in Heterogeneous Networks.}
\newblock \bibinfo{journal}{\emph{CoRR}}  \bibinfo{volume}{abs/1309.7393}
  (\bibinfo{year}{2013}).
\newblock
\urldef\tempurl%
\url{http://dblp.uni-trier.de/db/journals/corr/corr1309.html#ShiKHYW13}
\showURL{%
\tempurl}


\bibitem[\protect\citeauthoryear{Sun, Han, Yan, Yu, and Wu}{Sun
  et~al\mbox{.}}{2011}]%
        {PathSim}
\bibfield{author}{\bibinfo{person}{Yizhou Sun}, \bibinfo{person}{Jiawei Han},
  \bibinfo{person}{Xifeng Yan}, \bibinfo{person}{Philip~S. Yu}, {and}
  \bibinfo{person}{Tianyi Wu}.} \bibinfo{year}{2011}\natexlab{}.
\newblock \showarticletitle{Pathsim: Meta path-based top-k similarity search in
  heterogeneous information networks}.
  \bibinfo{howpublished}{\url{https://bit.ly/2Hx6XTM}}. In
  \bibinfo{booktitle}{\emph{In VLDB 11}}.
\newblock


\bibitem[\protect\citeauthoryear{Tavakolifard, Gulla, Almeroth, Ingvaldesn,
  Nygreen, and Berg}{Tavakolifard et~al\mbox{.}}{2013}]%
        {Tavakolifard:2013:TNP:2487788.2487930}
\bibfield{author}{\bibinfo{person}{Mozhgan Tavakolifard},
  \bibinfo{person}{Jon~Atle Gulla}, \bibinfo{person}{Kevin~C. Almeroth},
  \bibinfo{person}{Jon~Espen Ingvaldesn}, \bibinfo{person}{Gaute Nygreen},
  {and} \bibinfo{person}{Erik Berg}.} \bibinfo{year}{2013}\natexlab{}.
\newblock \showarticletitle{Tailored News in the Palm of Your Hand: A
  Multi-perspective Transparent Approach to News Recommendation}. In
  \bibinfo{booktitle}{\emph{Proceedings of the 22Nd International Conference on
  World Wide Web}} \emph{(\bibinfo{series}{WWW '13 Companion})}.
  \bibinfo{publisher}{ACM}, \bibinfo{address}{New York, NY, USA},
  \bibinfo{pages}{305--308}.
\newblock
\showISBNx{978-1-4503-2038-2}
\urldef\tempurl%
\url{https://doi.org/10.1145/2487788.2487930}
\showDOI{\tempurl}


\bibitem[\protect\citeauthoryear{Wen, Fang, and Guan}{Wen
  et~al\mbox{.}}{2012}]%
        {Wen:2012:HAP:2109236.2109445}
\bibfield{author}{\bibinfo{person}{Hao Wen}, \bibinfo{person}{Liping Fang},
  {and} \bibinfo{person}{Ling Guan}.} \bibinfo{year}{2012}\natexlab{}.
\newblock \showarticletitle{A Hybrid Approach for Personalized Recommendation
  of News on the Web}.
\newblock \bibinfo{journal}{\emph{Expert Syst. Appl.}} \bibinfo{volume}{39},
  \bibinfo{number}{5} (\bibinfo{date}{April} \bibinfo{year}{2012}),
  \bibinfo{pages}{5806--5814}.
\newblock
\showISSN{0957-4174}
\urldef\tempurl%
\url{https://doi.org/10.1016/j.eswa.2011.11.087}
\showDOI{\tempurl}


\bibitem[\protect\citeauthoryear{Xu, Jiang, and Watcharawittayakul}{Xu
  et~al\mbox{.}}{2017}]%
        {xu-jiang-watcharawittayakul:2017:Long}
\bibfield{author}{\bibinfo{person}{Mingbin Xu}, \bibinfo{person}{Hui Jiang},
  {and} \bibinfo{person}{Sedtawut Watcharawittayakul}.}
  \bibinfo{year}{2017}\natexlab{}.
\newblock \showarticletitle{A Local Detection Approach for Named Entity
  Recognition and Mention Detection}. In \bibinfo{booktitle}{\emph{Proceedings
  of the 55th Annual Meeting of the Association for Computational Linguistics
  (Volume 1: Long Papers)}}. \bibinfo{publisher}{Association for Computational
  Linguistics}, \bibinfo{address}{Vancouver, Canada},
  \bibinfo{pages}{1237--1247}.
\newblock
\urldef\tempurl%
\url{http://aclweb.org/anthology/P17-1114}
\showURL{%
\tempurl}


\bibitem[\protect\citeauthoryear{Yang and Yeung}{Yang and Yeung}{2010}]%
        {a4e1b868beda4b46a6723ba3d024ec41}
\bibfield{author}{\bibinfo{person}{Linda Yang} {and} \bibinfo{person}{K.
  Yeung}.} \bibinfo{year}{2010}\natexlab{}.
\newblock \bibinfo{title}{A proactive personalized mobile news recommendation
  system}.
\newblock
\newblock


\bibitem[\protect\citeauthoryear{Zhu and Iglesias}{Zhu and Iglesias}{2017}]%
        {7572993}
\bibfield{author}{\bibinfo{person}{G. Zhu} {and} \bibinfo{person}{C.~A.
  Iglesias}.} \bibinfo{year}{2017}\natexlab{}.
\newblock \showarticletitle{Computing Semantic Similarity of Concepts in
  Knowledge Graphs}.
\newblock \bibinfo{journal}{\emph{IEEE Transactions on Knowledge and Data
  Engineering}} \bibinfo{volume}{29}, \bibinfo{number}{1} (\bibinfo{date}{Jan}
  \bibinfo{year}{2017}), \bibinfo{pages}{72--85}.
\newblock
\showISSN{1041-4347}
\urldef\tempurl%
\url{https://doi.org/10.1109/TKDE.2016.2610428}
\showDOI{\tempurl}


\end{thebibliography}

\thanks{This work was partially supported by the Natural Sciences and Engineering Research Council of Canada, under a USRA  grant No. 511876.}

\clearpage

\end{document}